\documentclass[pre,amsmath,amssymb,notitlepage,showpacs]{revtex4-1}

\usepackage{epsfig}

\def\be{{\bf e}}
\def\bv{{\bf v}}
\def\bu{{\bf u}}
\def\br{{\bf r}}
\def\bx{{\bf x}}
\def\bk{{\bf k}}

\def\mellin{\mathcal{C}}
\def\hydro{\mathcal{H}}

\newcommand{\rcite}[1]{Ref.~\cite{#1}}
\newcommand{\rcites}[1]{Refs.~\cite{#1}}
\newcommand{\eq}[1]{Eq.~(\ref{#1})}
\newcommand{\eqs}[1]{Eqs.~(\ref{#1})}
\newcommand{\fig}[1]{Fig.~\ref{#1}}

\def\pnas{{Proc.\ Nat.\ Acad.\ Sci.}}

\begin{document}

\title{Signature of the time--dependent hydrodynamic interactions \\
  on the collective diffusion in colloidal monolayers}

\author{Alvaro Dom{\'i}nguez}

\affiliation{F\'\i sica Te\'orica, Universidad de Sevilla, Apdo.~1065,
  41080 Sevilla, Spain}\email{dominguez@us.es}

\date{October 9, 2014}

\begin{abstract}

  It has been shown recently that the coefficient of collective
  diffusion in a colloidal monolayer is divergent due to the
  hydrodynamic interactions mediated by the ambient fluid in bulk.
  The analysis is extended to allow for time--dependent hydrodynamic
  interactions. Novel observational features specific to this time
  dependency are predicted. The possible experimental detection in the
  dynamics of the monolayer is discussed.

\end{abstract}



\maketitle

\section{Introduction}

A colloidal monolayer is formed when colloidal particles are
constrained to stay in a surface. This confinement can be achieved in
several manners:
particles trapped by wetting forces at the interface between two
fluids (typically, air and water, or oil and water) \cite{Bink02},
non-buoyant particles sedimented at the bottom of a fluid phase
\cite{ZLM99}, particles trapped by optical tweezers into predetermined
configurations \cite{LKB04}. The monolayer behaves for most practical
purposes as a two--dimensional (2D) system. This renders the monolayer
a practical physical system to address fundamental questions
experimentally that concern the role of the spatial dimensionality on
the mechano-statistical properties of many--body systems. A nice
illustration of this usefulness was the first experimental
confirmation of the Kosterlitz--Thouless scenario for melting in 2D
systems \cite{ZLM99}.

The dynamics of colloids may be strongly influenced by the hydrodynamic
interactions mediated by the ambient fluid in which they are immersed
(see, e.g., \rcite{Dhon96}). Therefore, although a colloidal monolayer
is a 2D system, the dynamics may include a contribution from 3D
hydrodynamic interactions by an unconfined ambient fluid. This
configuration can be called \textit{``partial confinement''} of the
system ``colloid + ambient fluid'', and it provides a different
scenario from absence of confinement (i.e., 3D colloid in bulk) and
from complete confinement (a monolayer embedded in a likewise confined
fluid, e.g., in a slit pore or in a liquid film \cite{DKIL08}).

Recently, the analysis of a theoretical model for the ``partial
confinement'' configuration has shown \cite{BDGH14} that, as a
consequence of the hydrodynamic interactions, the collective diffusion
in the monolayer is anomalous 
on spatial scales above a certain characteristic length
$L_\mathrm{hydro}$. This prediction has been confirmed experimentally
by the measurement, through dynamic light scattering, of a diverging
coefficient of collective diffusion \cite{LRW95,LCXZ14}.
This unique feature of ``partial confinement'' (as opposed to absence
of and to complete confinement) follows from the contribution of the
long--ranged part of the hydrodynamic interactions in the linearized
equation for density perturbations.
This theoretical prediction assumes that the hydrodynamic interactions
are established instantaneously. This is a good approximation in many
experimental situations and simplifies considerably the theoretical
modelling.  However, there can be configurations in which the
relaxation of the ambient flow vorticity cannot be neglected. In this
work, the model introduced in \rcite{BDGH14} is extended to
incorporate this effect. The analysis reveals a second characteristic
length $L_\mathrm{cross}$ (much larger than $L_\mathrm{hydro}$), above
which the dynamics of density perturbations crosses over to a
non--diffusive behavior dominated by the time--dependency of the
hydrodynamic interactions, that exhibits specific features associated
to the ``partial confinement''.
This is actually a novel prediction for the observation of
time--dependent hydrodynamic interactions, facilitated in this case by
the configuration of ``partial confinement''.

In Sec.~\ref{sec:theory} we introduce the extension of the theoretical
model of \rcite{BDGH14} that incorporates the evolution of the
velocity field in the ambient flow, modelled with the time--dependent
Stokes equation. An equation for the evolution of the monolayer
density is derived and solved in the limit of small deviations from
homogeneity. Two opposite limiting cases of the solution are
discussed, namely, the limit of instantaneous establishment of the
hydrodynamic interaction (Sec.~\ref{sec:notimeHI}), so that the model
of \rcite{BDGH14} is recovered, and the ``thermodynamic limit'' of
density perturbations with infinite spatial extension
(Sec.~\ref{sec:timeHI}), so that the crossover length scale
$L_\mathrm{cross}$ is probed by the dynamics. In Sec.~\ref{sec:exp}
the possibility of the experimental observation of time--dependent
hydrodynamic interactions via the monolayer dynamics is thoroughly
discussed and estimates are provided for the relevant length and time
scales in realistic experimental conditions. Sec.~\ref{sec:end}
summarizes the conclusions.

\section{Theoretical model}
\label{sec:theory}

We consider the simplest physical model that exhibits the relevant
phenomenology as introduced in \rcite{BDGH14}, namely a collection of
particles restricted to move in the plane $z=0$ but subjected to the
hydrodynamic interaction mediated by an ambient fluid filling the
whole space. As discussed in \rcite{BDO14}, modifications of this
model to describe more realistic experimental configurations (e.g.,
particles confined to the planar interface at $z=0$ between two
different fluids) do not alter the qualitative picture. The following
fields are defined (with $\br =
(x,y)$ denoting the position in the monolayer plane at $z=0$): \\
(i) The 2D particle number density field, $\varrho (\br, t)$, in the
monolayer plane. \\
(ii) The in--plane (2D) velocity field, $\bv(\br, t)$, of the flow of
particles. \\
(iii) The 3D velocity field, $\bu(\br, z, t)$, of the flow of the ambient fluid. \\
(iv) The average total force per particle,
$\mathbf{f}_\mathrm{tot}(\br, t)$, which is a ``generalized'' or
``thermodynamic'' force and accounts for the effect of Brownian
diffusion, of the direct (``static'') interactions between the
particles, and of the external and
confining forces. \\
These fields are related to each other by the following equations,
expressing physical laws and simplifying assumptions:

\noindent
\textbf{(A)} Particle number conservation in the monolayer plane:
  \begin{equation}
    \label{eq:cont}
    \frac{\partial\varrho}{\partial t} = - \nabla\cdot (\varrho \bv ) ,
    \quad
    \nabla := \left( \frac{\partial}{\partial x}, \frac{\partial}{\partial y} \right).
  \end{equation}

\noindent
\textbf{(B)} Particle motion
  in the so-called \textit{point--particle (Oseen) approximation}:
  \begin{equation}
    \label{eq:vfield}
    \bv(\br, t) = \Gamma \mathbf{f}_\mathrm{tot}(\br, t) + \bu(\br, z=0, t) ,
  \end{equation}
  together with the time--dependent Stokes equation for incompressible
  flow (denoting $\bx := \br + z \be_z$ and $\nabla_\bx := \nabla +
  \be_z \partial_z$),
  \begin{subequations}
    \label{eq:stokes}
    \begin{equation}
      \label{eq:dynstokes}
      \rho_\mathrm{fluid} \frac{\partial \bu}{\partial t} = 
      \eta \nabla_\bx^2 \bu - \nabla_\bx p + \delta(z) \varrho(\br, t) 
      \mathbf{f}_\mathrm{tot} (\br,t) ,
    \end{equation}
    \begin{equation}
      \label{eq:incomp1}
      \nabla_\bx\cdot\bu = 0 ,
    \end{equation}
  \end{subequations}
  with the boundary condition of vanishing fields at infinity. Here,
  $\Gamma$ is the mobility of an isolated particle,
  $\rho_\mathrm{fluid}$ is the mass density of the ambient fluid,
  $\eta$ its kinematic viscosity, $p(\bx)$ is the pressure field
  enforcing the 3D incompressibility constraint~(\ref{eq:incomp1}),
  and the Dirac delta in \eq{eq:dynstokes} describes the geometrical
  confinement of the particles to the plane $z=0$.
  Physically, \eq{eq:vfield} represents the motion of a particle in
  the overdamped regime under the effect of the total force as if
  isolated (Stokes drag) plus the advection by the ambient fluid flow,
  \textit{evaluated at the confining plane}. The ambient flow, in
  turn, is determined by \eqs{eq:stokes} self--consistently in terms of
  the motion of the particles. 

  The point--particle approximation incorporates only the dominant
  contribution of the hydrodynamic interaction in the limit that the
  interparticle separation is much larger than the size and the
  hydrodynamic radius of the particles (dilute regime). It is possible
  to relax this hypothesis to some extent by allowing for a dependence
  of $\Gamma$ on the density $\varrho$, which should account for the
  short--separation contributions by the hydrodynamic interaction
  (see, e.g., \rcites{Batc72,BrDu88,ToOp94,ToOp95,HaIc95,MLP10}). This
  would not affect, however, the conclusions \cite{BDO14}.

  This model for the ambient flow includes the diffusion of the
  vorticity in the regime of low Reynolds and Mach numbers. This is
  the point of departure from the model addressed in
  \rcites{BDGH14,BDO14}, which assumes that the ambient flow adapts
  instantaneously to a given particle configuration (i.e., one sets
  $\partial_t \bu = 0$ in \eq{eq:dynstokes}).

\noindent
\textbf{(C)} Particles are confined to the plane $z=0$, i.e., $\be_z \cdot
\bv(\br, t) = 0$, implying from \eq{eq:vfield} that
  \begin{equation}
    \label{eq:novz}
    \be_z \cdot \mathbf{f}_\mathrm{tot} (\br, t) 
    = - \frac{1}{\Gamma} \be_z \cdot \bu (\br, z=0,t) .    
  \end{equation}
  It is straightforward to show that
  if the pair $\{\bu(\br,z,t), \mathbf{f}_\mathrm{tot}(\br,t)\}$
  satisfies \eqs{eq:stokes}, so does the specular reflection with
  respect to the plane $z=0$ (i.e., the pair obtained by the
  transformation $z\to -z$, $u_z\to -u_z$,
  $\be_z\cdot\mathbf{f}_\mathrm{tot}\to -
  \be_z\cdot\mathbf{f}_\mathrm{tot}$, everything else unchanged). 
  Therefore, a solution to \eq{eq:novz} is
  \begin{equation}
    \label{eq:nofz}
    \be_z \cdot \mathbf{f}_\mathrm{tot} (\br, t) = 0    
  \end{equation}
  by continuity of the velocity field $\bu$ at $z=0$. Physically, if
  the net force on the particles points in the confining plane, their
  motion only induces, in the point--particle approximation, an
  in-plane ambient flow when evaluated at the plane.

  The condition expressed by \eq{eq:nofz} is actually an equation for
  the unknown constraining force, that can be used to eliminate any
  explicit mention to this force in the model equations: one can
  replace $\mathbf{f}_\mathrm{tot}(\br,t)$ in
  Eqs.~(\ref{eq:vfield},\ref{eq:stokes}) by the projection onto the
  $z=0$ plane of all the forces other than the constraining force
  (i.e., Brownian, interparticle, and external). We denote this
  projection simply as $\mathbf{f}(\br,t)$, which by construction
  points in the confining plane.

\noindent
\textbf{(D)} This latter force $\mathbf{f}(\br, t)$ is assumed to be given as
  a function solely of the density field $\varrho(\br, t)$. A usual
  implementation of this assumption is the approximation of local
  thermal equilibrium, i.e., at each point the colloidal monolayer is
  assumed locally in intrinsic local equilibrium, and the flow of
  particles is driven by gradients of the local chemical potential as
  given by thermodynamics (and thus given as a function of the local
  density at the isothermal conditions appropriate for a colloid):
  \begin{equation}
    \label{eq:localf}
    \mathbf{f} = - \nabla \mu 
    = - \left.\frac{\partial \mu}{\partial \varrho}\right)_T
    \nabla \varrho .
  \end{equation}
  More sophiscated approximations to the functional form of
  $\mathbf{f}$ can be used in order to get expressions valid in a
  wider range of length scales or in situations very far from
  equilibrium. However, here only \eq{eq:localf} will be used for
  simplicity, because it suffices to illustrate the phenomenology we
  are interested in.

  In conclusion, Eqs.~(\ref{eq:cont}--\ref{eq:localf}) form a closed
  set of equations for the evolution of the monolayer as described by
  the density field $\varrho(\br,t)$.
Assuming that there are no external force fields, $\mathbf{f}(\br)$
will vanish in a homogeneous state, $\varrho(\br, t) =
\varrho_\mathrm{hom}$, so that the latter is a stationary solution of
the model equations. They can be linearized about this reference
solution, $\varrho(\br, t) = \varrho_\mathrm{hom} + \delta
\varrho(\br, t)$ with $|\delta \varrho| \to 0$, so that
\begin{subequations}
  \label{eq:lin}
  \begin{equation}
    \label{eq:lincont}
    \frac{\partial \delta\varrho}{\partial t} \approx 
    - \Gamma \varrho_\mathrm{hom} \nabla\cdot\delta\mathbf{f} 
    - \varrho_\mathrm{hom} \nabla\cdot\bu(\br,z=0) ,
  \end{equation}
  \begin{equation}
    \label{eq:linstokes}
    \rho_\mathrm{fluid} \frac{\partial \bu}{\partial t} \approx 
    \eta \nabla_\bx^2 \bu - \nabla_\bx p + \delta(z) \varrho_\mathrm{hom}
    \delta\mathbf{f} (\br,t) ,
  \end{equation}
  \begin{equation}
    \label{eq:incomp2}
    \nabla_\bx\cdot\bu = 0 ,      
  \end{equation}
  \begin{equation}
    \delta \mathbf{f} \approx - \frac{D_0}{\Gamma\varrho_\mathrm{hom}}
    \nabla \delta \varrho ,    
  \end{equation}
  \begin{equation}
    \label{eq:D0}
    D_0 := \Gamma \varrho_\mathrm{hom} 
    \left.\frac{\partial \mu}{\partial \varrho}\right)_T (\varrho=\varrho_\mathrm{hom}) .
  \end{equation}
\end{subequations}
The coefficient of collective diffusion, $D_0$, can be related to the
isothermal compressibility of the monolayer in the reference
homogeneous state (see \eq{eq:D0comp}).
Although one usually considers $D_0>0$ (the reference homogeneous
state is stable), it is also of interest to consider the influence of
the hydrodynamic interactions on the dynamics of an unstable state,
$D_0<0$ (an example of experimental relevance is the clustering in a
monolayer under the effect of capillary attraction
\cite{DOD10,BDDO11,BDOD11,BDGH14,BDOD14}). Therefore, in the calculations in the
rest of the paper no assumption will be made concerning the sign of
$D_0$.

It is instructive to discuss how \eqs{eq:lin} would be modified in the
cases of absence of or complete confinement, respectively. The key issue
is that, although the 3D ambient flow is incompressible, see
\eq{eq:incomp2}, the continuity equation for the monolayer
incorporates only the ambient flow \textit{in the monolayer plane,
  $z=0$}, see \eq{eq:lincont}, and this needs not be incompressible.
This fact and the long--ranged nature of the hydrodynamic interactions
co-act to yield a divergent correction of the coefficient of
collective diffusion in the limit of instantaneous hydrodynamic
interactions \cite{BDGH14}.
In the absence of confinement (i.e., 3D continuity equation in 3D
ambient flow) or in the case of complete confinement (2D continuity
equation in 2D ambient flow), there would be no explicit dependence on
the ambient velocity field $\bu$ in the linearized continuity equation
due to the incompressibility constraint. This does not mean that
hydrodynamic interactions would not affect diffusion, but only that
this would occur through nonlinear corrections or, more generally,
through mode--coupling terms that would also incorporate the
short--range contributions by the hydrodynamic interactions. Explicit
calculations and simulations of a 3D colloid (see, e.g.,
\rcites{Batc72,BrDu88,ToOp94,ToOp95,HaIc95,MLP10}) show that these
effects lead at most to a finite renormalization of the diffusion
coefficient in the form of a density--dependent mobility, as mentioned
briefly before.

In order to solve \eqs{eq:lin}, one introduces the Laplace transform
in time and the Fourier transform in the spatial variables: 
\begin{subequations}
\begin{equation}
  \hat{\varrho}(\bk,s) := \int_0^{+\infty} dt\; \mathrm{e}^{-s t} \int d^2 \br\; 
  \mathrm{e}^{-i \bk\cdot\br} \delta\varrho(\br, t) ,
\end{equation}
\begin{equation}
  \hat{\bu}(\bk,q,s) := \int_0^{+\infty} dt\; \mathrm{e}^{-s t} 
  \int d^2 \br\; \int_{-\infty}^{+\infty} dz\;
  \mathrm{e}^{-i \bk\cdot\br - i q z} \bu(\br, z, t) .
\end{equation}
\end{subequations}
For an initial condition $\varrho (\br, t=0) = \varrho_0(\br)$,
$\bu(\br,z,t=0) = \mathbf{0}$, \eqs{eq:lin} become
\begin{subequations}
  \label{eq:lintransf}
  \begin{equation}
    \label{eq:conttransf}
    s \hat{\varrho}(\bk, s) - \hat{\varrho}_0(\bk) \approx - D_0 k^2 \hat{\varrho}(\bk, s)
    - \varrho_\mathrm{hom} 
    \int_{-\infty}^{+\infty} \frac{dq}{2\pi}\; i\bk\cdot \hat{\bu}(\bk, q, s) ,
  \end{equation}
  \begin{equation}
    \label{eq:stokestransf}
      \left[ \rho_\mathrm{fluid} s + \eta (k^2+q^2) \right] \hat{\bu}(\bk, q, s)
    \approx - \frac{D_0}{\Gamma} \left[ 
      \mathcal{I} - \frac{(\bk+q\be_z) (\bk+q\be_z)}{k^2+q^2} 
    \right] \cdot (i\bk) \hat{\varrho}(\bk, s) . 
  \end{equation}
\end{subequations}
One can evaluate the integral term in \eq{eq:conttransf}:
\begin{eqnarray}
  \int_{-\infty}^{+\infty} \frac{dq}{2\pi}\; i\bk\cdot \hat{\bu}(\bk, q, s) 
  & = & 
  \frac{D_0 k^2}{\Gamma} \hat{\varrho}(\bk,s) \int_{-\infty}^{+\infty} \frac{dq}{2\pi}\; 
  \frac{q^2}{[k^2+q^2]\left[ \rho_\mathrm{fluid} s + \eta (k^2+q^2) \right]}
  \nonumber \\ 
  & = & \frac{D_0 k^2 \hat{\varrho}(\bk,s)}{\varrho_\mathrm{hom}} \mathcal{H}(k,s) ,
\end{eqnarray}
where we have defined the auxiliary function
\begin{equation}
  \label{eq:H}
  \hydro (k, s) := \frac{2}{L_\mathrm{hydro} k} \left[ 
    1 + \sqrt{1 + \frac{\tau_\omega s}{(L_\mathrm{hydro} k)^2} } \right]^{-1} ,
\end{equation}
in terms of the length scale
\begin{equation}
  \label{eq:Lhydro}
  L_\mathrm{hydro} := \frac{4\eta\Gamma}{\varrho_\mathrm{hom}} ,
\end{equation}
introduced in \rcite{BDGH14}, and the time scale
\begin{equation}
  \tau_\omega := \frac{\rho_\mathrm{fluid} L_\mathrm{hydro}^2}{\eta} ,
\end{equation}
associated to the relaxation of the ambient vorticity on the length
scale $L_\mathrm{hydro}$. The solution of \eq{eq:conttransf} is written as
\begin{equation}
  \label{eq:linrho}
  \hat{\varrho}(\bk, s) = \hat{\varrho}_0(\bk) \hat{G}(k, s),  
\end{equation}
with the Green function
\begin{equation}
  \label{eq:green}
  \hat{G}(k, s) := \frac{1}{s + D_0 k^2 [ 1 + \hydro (k, s) ]} .
\end{equation}
Thus, the function $\hydro (k,s)$ encodes the effect of the
hydrodynamic interactions on the dynamics of the monolayer.

The goal is to study the analytical properties of the Green function
$\hat{G}(k,s)$ and, more interestingly, of its inverse Laplace transform,
given by the Mellin formula,
\begin{subequations}
  \label{eq:mellin}
\begin{equation}
  G (k, t>0) = \int_{\mellin}
  \frac{ds}{2\pi i} \; \mathrm{e}^{s t} \hat{G}(k, s) ,
\end{equation}
with the integration path 
\begin{eqnarray}
  \label{eq:path}
  \mellin & := & \left\{ \mathrm{Re} \; s = p > 0 
    \textrm{ constant and to the right } \right. \\ 
  & & \left. \textrm{of any singularity in the complex $s$--plane} 
  \right\} . \nonumber
\end{eqnarray}
\end{subequations}
As we shall see, $G(k,t)$ is directly related to the experimentally
relevant intermediate scattering function. Its dependence on $t$
is controlled by the structure in the complex $s$--plane of the
function $\hat{G}(k,s)$. This is analyzed in App.~\ref{sec:complexG}
and here we summarize the relevant conclusions. A crossover length
scale, associated to the time scale $\tau_\omega$, appears naturally
as
\begin{equation}
  \label{eq:Lcross}
  L_\mathrm{cross} := \frac{L_\mathrm{hydro}^3}{\tau_\omega |D_0|} .
\end{equation}
Notice that, unlike $L_\mathrm{hydro}$, this length scale depends on
the specific form of the interaction potential between the particles
through the value of the diffusion coefficient $D_0$. It can be
assumed that $L_\mathrm{hydro} < L_\mathrm{cross}$ (actually,
$L_\mathrm{hydro} \ll L_\mathrm{cross}$ in realistic configurations,
see the discussion in Sec.~\ref{sec:exp}).
In the complex $s$--plane (see left part of \fig{fig:complexG}), the
function $\hat{G}(k,s)$ has a branch cut discontinuity in the negative
real axis with a branching point at 
\begin{equation}
  \label{eq:sbranch}
  s_\mathrm{branch} = - \frac{(L_\mathrm{hydro} k)^2}{\tau_\omega},
\end{equation}
and either (i) a single real pole if $D_0<0$, or if $D_0>0$ \emph{and}
$L_\mathrm{cross} k \gtrsim 1$, or (ii) two complex conjugate poles if
$D_0>0$ \emph{and} $L_\mathrm{cross} k \lesssim 1$.
Consequently, the inversion in \eq{eq:mellin} is written as the sum of
a contribution by the poles and a contribution by an integral along
the branch discontinuity, see \eq{eq:mellin3}. Although this general
expression can be applied to study the two cases, it is physically
more illuminating to consider two limiting situations which provide
the correct qualitative picture, namely, the limit
$L_\mathrm{cross}\to \infty$ for case (i), interpreted as
$\tau_\omega\to 0$ or time--independent hydrodynamic interactions, and
the limit $k\to 0$ for case (ii), interpreted as the ``thermodynamic
limit''.

\subsection{Time--independent hydrodynamic interactions}
\label{sec:notimeHI}

In the limiting case in which the hydrodynamic interactions are
established instantaneously, one recovers the results presented
in \rcite{BDGH14}. This limit means that the time $\tau_\omega$ is
much shorter than any other time of interest and corresponds
mathematically to the limit $\tau_\omega\to 0$ in the Green function:
from \eq{eq:H}, one gets $\hydro (k,s) \sim \hydro (k,0) =
1/(L_\mathrm{hydro} k)$ and from \eq{eq:green},
\begin{equation}
  \label{eq:greennotime}
  \hat{G}(k, s) \sim \frac{1}{s + D_0 k^2 [ 1 + \hydro (k,0) ]} .  
\end{equation}
It is useful to introduce the $k$--dependent time scales
\begin{equation}
  \label{eq:tau0}
  \tau_0 (k) := \frac{1}{|D_0| k^2}
\end{equation}
(corresponding to normal diffusion) and 
\begin{equation}
  \label{eq:tau1}
  \tau_1 (k) := \frac{\tau_0}{\hydro (k,0)} = \frac{L_\mathrm{hydro}}{|D_0| k} .
\end{equation}
Therefore, since the Green function in \eq{eq:greennotime} has only a
simple pole in $s$, \eq{eq:mellin} predicts an exponential dependence
in time,
\begin{equation}
  \label{eq:greendiff}
  G(k, t) = \mathrm{e}^{- (\mathrm{sign}\; D_0) t \left[ 
      \frac{1}{\tau_0} + \frac{1}{\tau_1} 
    \right]
    }.
\end{equation}
(That is, the inversion of the Laplace transform is dominated by the
contribution of the single pole in the real axis; the contribution of
the branch cut discontinuity is negligible because the branching point
moves to infinity in the limit $\tau_\omega \to 0$, see \eq{eq:sbranch}).
The meaning of the length scale $L_\mathrm{hydro}$ defined in
\eq{eq:Lhydro} is now clear: if $L_\mathrm{hydro} k \gg 1$, the
evolution of the Fourier modes is controlled by the time scale
$\tau_0(k)$ and they follow normal diffusion with a constant $D_0$
(actually, ``antidiffusion'' if $D_0<0$).
In the opposite case, $L_\mathrm{hydro} k \ll 1$ and the evolution is
controlled by the time scale $\tau_1(k)$, so that the Fourier modes
exhibit anomalous diffusion (superdiffusion, to be more precise),
i.e., exponential dependence in time with a diffusion coefficient
diverging as $k\to 0$. Physically, the length scale $L_\mathrm{hydro}$
separates the regimes when the evolution of the monolayer density is
dominated, in \eq{eq:lincont} or in \eq{eq:conttransf}, by diffusion
properly (and one approximates $1+\hydro (k,0) \approx 1$ in
\eq{eq:greennotime}) or by advection by the time--independent ambient
flow (and $1+\hydro (k,0) \approx \hydro(k,0)$ in
\eq{eq:greennotime}).

It is clear from \eq{eq:H} that the approximation $\tau_\omega\to 0$
and the ``thermodynamic'' limit, $k\to 0$, do not
commute.
Since the inverse Laplace transform is controlled by the pole at
$s=-\mathrm{sign}\; D_0/\tau_1(k)$ when setting $\tau_\omega=0$, the
previous conclusions are valid provided $k$ is small but still large
enough that it holds
\begin{equation}
  \label{eq:smalltauomega}
  \left.\frac{\tau_\omega |s|}{(L_\mathrm{hydro} k)^2} \right|_{s=1/\tau_1(k)} 
  \ll 1
  \quad\Leftrightarrow\quad 1 \ll L_\mathrm{cross} k ,
\end{equation}
in terms of the crossover length scale defined by \eq{eq:Lcross}.
Therefore, the regime of anomalous diffusion described by
\eq{eq:greendiff} must be interpreted as an intermediate asymptotics,
$L_\mathrm{hydro} \ll k^{-1} \ll L_\mathrm{cross}$. Physically, the
restriction in \eq{eq:smalltauomega} means that the ambient vorticity
of mode $k$ relaxes much faster than the characteristic time scale of
particle diffusion over a length scale $\sim k^{-1}$. If this scale is
so large that $L_\mathrm{cross} k\ll 1$, however, one cannot neglect
the dynamical evolution of the ambient fluid vorticity.  This is
addressed next.

\subsection{Time--dependent hydrodynamic interactions}
\label{sec:timeHI}

Consider now the ``thermodynamic limit'', i.e., the limit $k\to 0$ of
the Green function without the restriction described by
\eq{eq:smalltauomega}. As shown in App.~\ref{sec:smallk}, both the
branch point and the singularities of $\hat{G}(k,s)$ approach zero as
$k\to 0$, whereby a new time scale appears naturally that is defined
as
\begin{equation}
  \label{eq:tau2}
  \tau_2(k) := \left( \frac{\tau_\omega}{4 |D_0|^2 k^4} \right)^{1/3} ,
\end{equation}
to be compared with $\tau_0(k)$ and $\tau_1(k)$ in Eqs.~(\ref{eq:tau0},
\ref{eq:tau1}).
%
The inversion in \eq{eq:mellin} is thus dominated by the behavior of
the Green function near $s=0$ in the limit $k\to 0$, so that one can
approximate $1 + \hydro (k,s) \sim 1 + \hydro (0,s) \sim \hydro (0,s)
= 2/\sqrt{\tau_\omega s}$ and
\begin{equation}
  \label{eq:greentimeHI}
  \hat{G}(k, s) \sim
  \frac{1}{s + D_0 k^2 \hydro(0,s)} .
\end{equation}
Physically, this means that the evolution of the monolayer density is
dominated, in \eq{eq:lincont} or in \eq{eq:conttransf}, by advection
by the ambient flow, in turn determined by neglecting the in--plane
shear in \eq{eq:linstokes} or in \eq{eq:stokestransf}, but not the time
dependence.

The evaluation of the Mellin formula (\ref{eq:mellin}) with the
approximation~(\ref{eq:greentimeHI}) is described in
App.~\ref{sec:smallk}. The result is summarized in the scaling behavior
\begin{subequations}
  \label{eq:greensmallk}
  \begin{equation}
    \label{eq:greensmallkscaling}
    G(k,t) = \gamma\left(\frac{t}{\tau_2(k)}\right) ,
  \end{equation}
  with the scaling function
  \begin{equation}
    \label{eq:gamma}
    \gamma(u) = \left\{
      \begin{array}[c]{cl}
        \displaystyle \frac{4}{3} \mathrm{e}^{-u/2} \cos \frac{\sqrt{3}}{2} u 
        - I(u) , &
        D_0 > 0 , \\ 
        & \\
        \displaystyle \frac{2}{3} \mathrm{e}^{u} 
        + I(u) , &
        D_0 < 0 ,
      \end{array} 
    \right.
  \end{equation}
  where \footnote{This function can actually be written in terms of a
    generalized hypergeometric function.}
  \begin{equation}
    \label{eq:I}
    I(u) := \frac{1}{\pi} \int_0^{\infty} dx\; \frac{\sqrt{x}}{x^3+1} \mathrm{e}^{-u x} .
  \end{equation}
\end{subequations}
Unlike in the limiting case addressed in the previous subsection, the
contribution of the branch cut is now as important as the one by the
singularities of the Green function.

\begin{figure}
  \hfill\epsfig{file=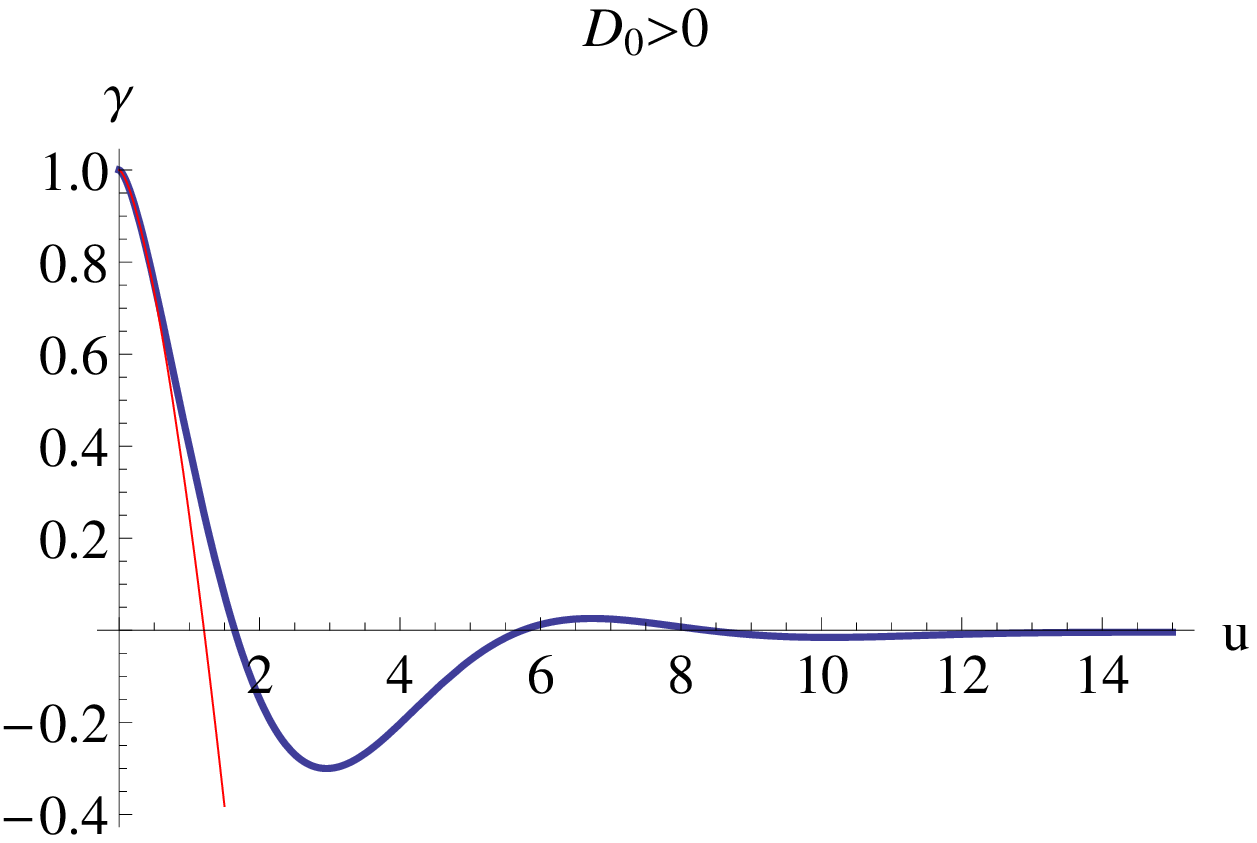,width=.4\textwidth}
  \hfill\epsfig{file=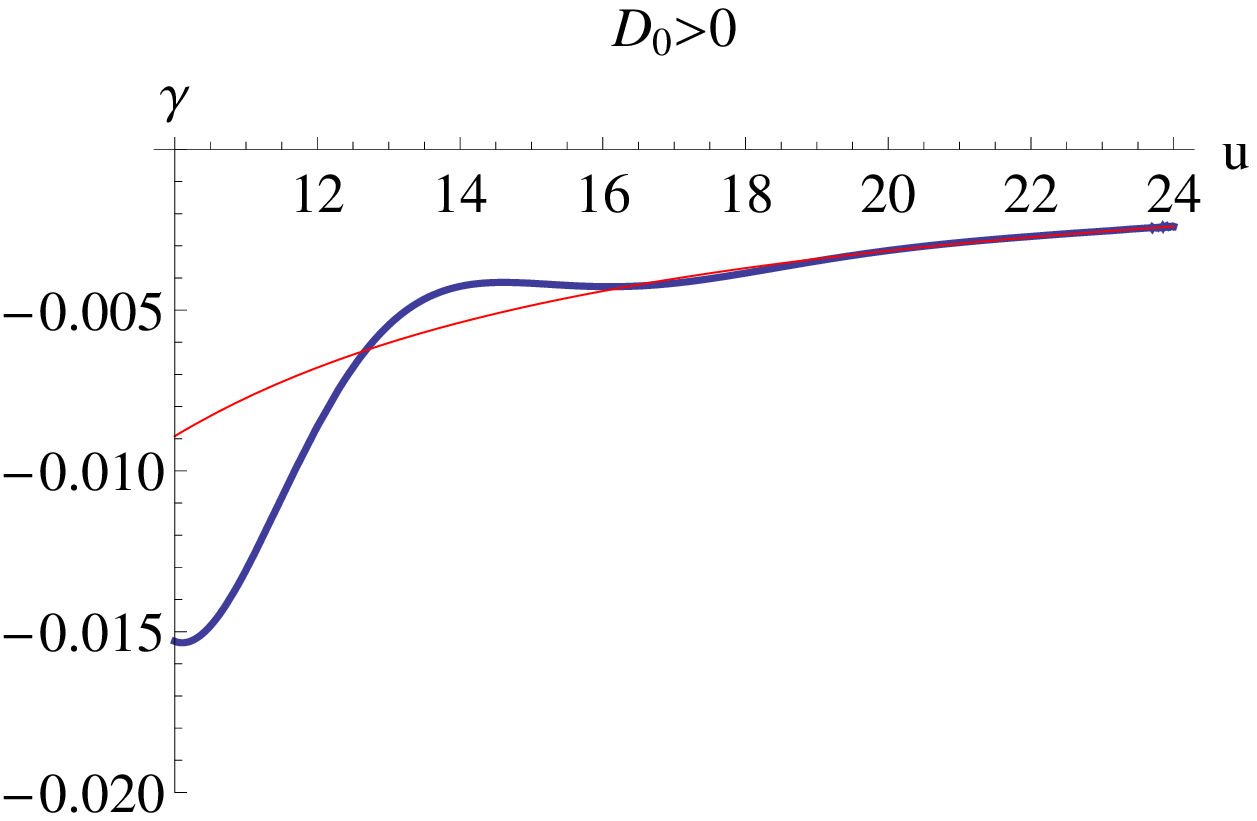,width=.4\textwidth}
  \hspace*{\fill}
  \caption{The scaling function $\gamma(u)$ for $D_0>0$ given by
    \eq{eq:gamma} (thick, blue line), together with the predicted
    asymptotic behaviors for small and large values of the argument,
    see Eqs.~(\ref{eq:smallu}) and~(\ref{eq:largeu}) respectively
    (thin, red lines).
    \label{fig:greenktstab}}
\end{figure}

\begin{figure}
  \hfill\epsfig{file=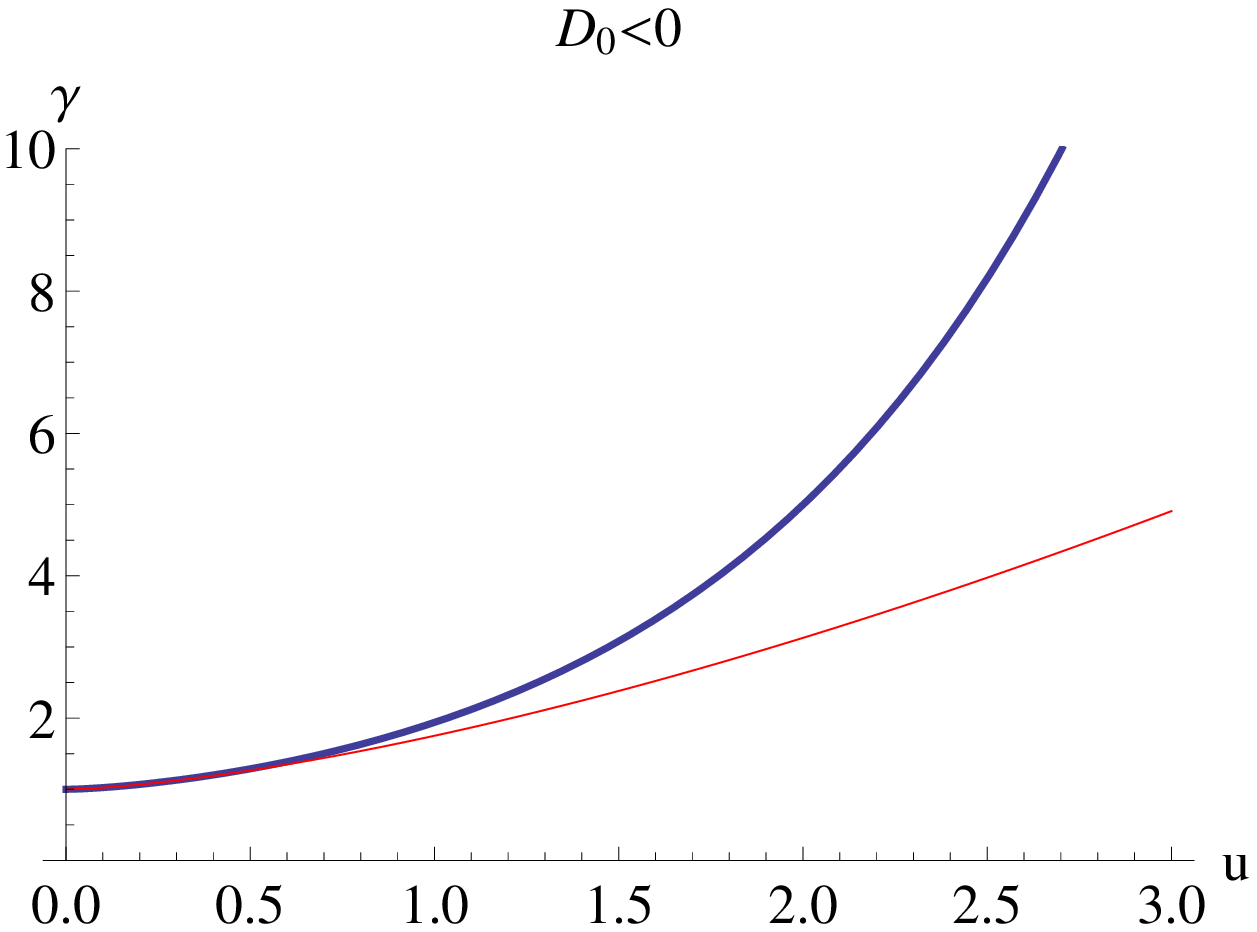,width=.4\textwidth}
  \hfill\epsfig{file=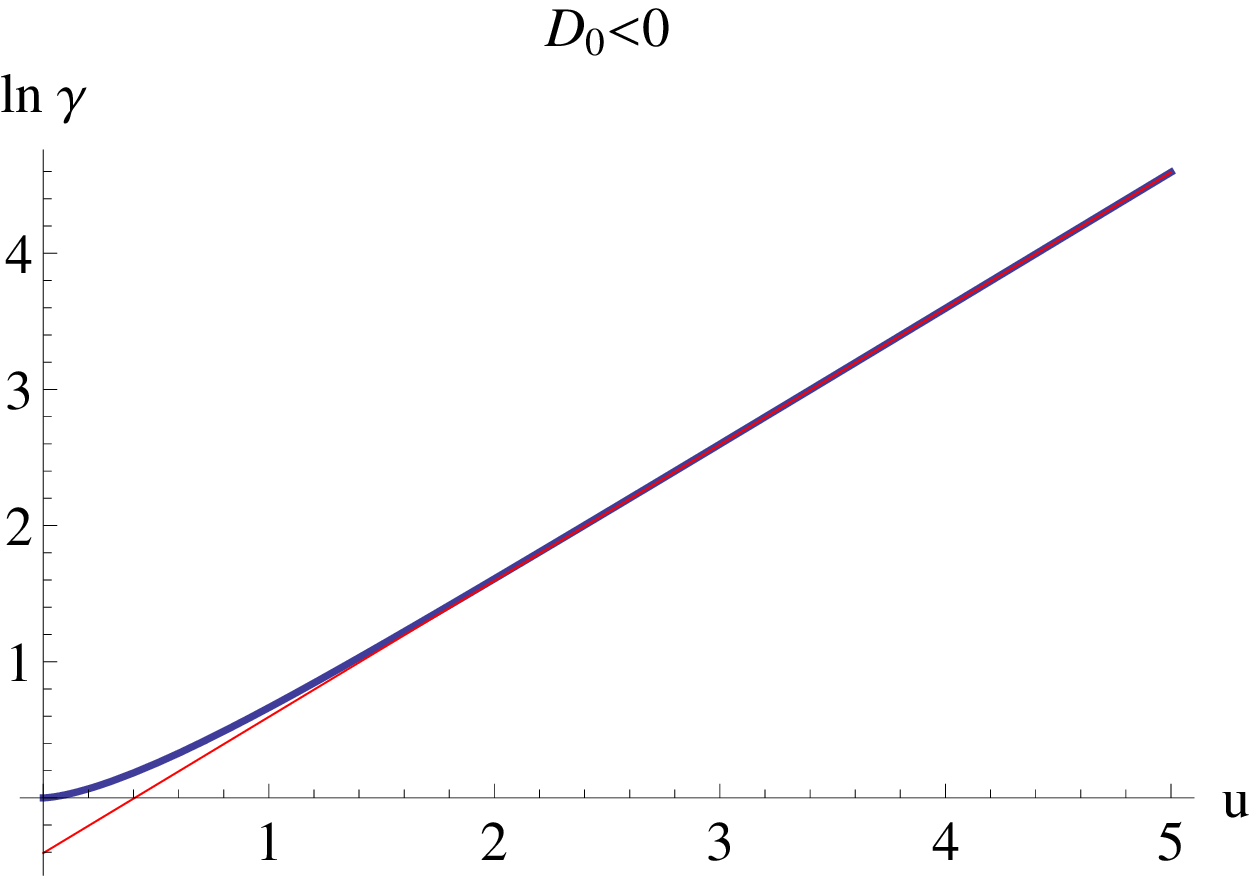,width=.4\textwidth}
  \hspace*{\fill}
  \caption{Same as \fig{fig:greenktstab} but for $D_0<0$.
    \label{fig:greenktunstab}}
\end{figure}

Figures~\ref{fig:greenktstab} and \ref{fig:greenktunstab} show plots
of the scaling function $\gamma(u)$.  Some relevant properties are
derived in App.~\ref{sec:greenkt}: the function $\gamma(u)$ is regular
and infinitely differentiable if $u>0$; at $u=0$, its second
derivative does not exist, behaving as
\begin{equation}
  \label{eq:smallu}
  \gamma(u\to 0) \sim 1 - \frac{4\mathrm{sign}\;D_0}{3\sqrt{\pi}} u^{3/2} .
\end{equation}
In the opposite limit one has the asymptotic behavior 
\begin{equation}
  \label{eq:largeu}
  \gamma(u\to+\infty)\sim \left\{
    \begin{array}[c]{cl}
      \displaystyle -\frac{1}{2\sqrt{\pi} u^{3/2}} , & 
      D_0 > 0 , \\
      & \\
      \displaystyle \frac{2}{3} \mathrm{e}^{u} , &
      D_0 < 0 .
    \end{array}
  \right.
\end{equation}
When $D_0>0$, the oscillatory behavior in time of $G(k,t)$ (with a
characteristic time $\tau_2(k)$) and the long--time algebraic decay,
$G(k,t) \sim t^{-3/2}$, are the signatures of the time dependence of
the hydrodynamic interactions.
If $D_0<0$, $G(k,t)$ grows exponentially in time, as in the absence of
hydrodynamic interactions, but the time dependence of the hydrodynamic
interaction shows up in the specific $k$--dependence of the
characteristic time scale, $\tau_2\propto k^{-4/3}$.

\section{Feasibility of experimental observation}
\label{sec:exp}

The Green function $G(k,t)$ has an immediate relationship with
experimentally accessible quantities because it is basically the
intermediate scattering function $F(k,t)$ of the monolayer, which can
be measured with dynamic light scattering experiments \cite{Dhon96}. If the
monolayer consists of $N$ particles, the function $F(k,t)$ is defined in terms of
an ensemble average over equilibrium configurations as \cite{HaMc86}
\begin{equation}
  \label{eq:scatt}
  F(k,t) := \frac{1}{N} \langle \hat{\varrho}(\bk,t) \hat{\varrho}(-\bk,0) \rangle 
  = G(k,t) S(k) ,
\end{equation}
when \eq{eq:linrho} is applied. Here, the structure factor $S(k) :=
\langle |\hat{\varrho}_0(\bk)|^2 \rangle/N$
is independent of the hydrodynamic interactions and determined
completely by thermodynamic equilibrium. If $k^{-1}$ is larger than
the equilibrium correlation length, one can approximate $S(k)$ by a
constant and the relevant dependence on $k$ and $t$ is determined
completely by the Green function.

The results of the previous Section can be summarized in the following
three expected regimes for the intermediate scattering function (for
definiteness, we assume the case of a stable reference state, i.e.,
$D_0>0$):
\begin{subequations}
  \label{eq:taularge}
  \\
  \textbf{(1)} \textit{Regime of negligible hydrodynamic interactions,
    $k^{-1} \ll L_\mathrm{hydro}$}: the function $F(k,t)$ exhibits
  normal diffusive behavior, i.e., exponential decay in time with a
  characteristic time that scales as $k^{-2}$, see \eq{eq:tau0}, so that it
  is possible to define a diffusion coefficient.
  \\
  \textbf{(2)} \textit{Regime of dominant instantaneous hydrodynamic
  interactions, $L_\mathrm{hydro} \ll k^{-1} \ll L_\mathrm{cross}$}:
  the function $F(k,t)$ still decays exponentially in time, but with a
  characteristic time that scales as $k^{-1}$, see \eq{eq:tau1}, and
  satisfying
  \begin{equation}
    \frac{\tau_1(k)}{\tau_\omega} = \frac{L_\mathrm{cross}}{L_\mathrm{hydro}} 
    \frac{1}{L_\mathrm{hydro} k}
    \gg 1 .
  \end{equation}
  Therefore, it is possible to define a $k$--dependent diffusion
  coefficient, but not a diffusion constant as the limit $k\to 0$.
  \\
  \textbf{(3)} \textit{Regime of dominant time--dependent hydrodynamic
    interactions, $L_\mathrm{cross} \ll k^{-1}$}: the function
  $F(k,t)$ exhibits damped oscillations and algebraic decay in time
  (see \fig{fig:greenktstab}) with a characteristic time that scales as
  $k^{-4/3}$, see \eq{eq:tau2}, and satisfying
  \begin{equation}
    \label{eq:tau2large}
    \frac{\tau_2(k)}{\tau_\omega} = \left(\frac{L_\mathrm{cross}}{L_\mathrm{hydro}}\right)^2 
    \frac{1}{(\sqrt{2} L_\mathrm{cross} k)^{4/3}}
    \gg 1 .
  \end{equation}
  Therefore, the dynamics of the monolayer density cannot be
  characterized as diffusive at all.
\end{subequations}

In order to provide quantitative estimates, we consider by
way of example a collection of spherical colloidal particles (radius
$R$) immersed in water at room temperature, for which $\eta \approx
10^{-3} \textrm{ N s/m$^2$}$, $\rho_\mathrm{fluid} \approx 10^{3}
\textrm{ kg/m$^3$}$. The 2D packing fraction of the colloid is denoted
as $\phi = \pi R^2 \varrho_\mathrm{hom}$. 
The particle mobility $\Gamma$ can depend on $\phi$ in order to
account for the effect of the short--distance hydrodynamic
interactions; this would amount to replacing $\Gamma$ by
$\Gamma(\varrho=\varrho_\mathrm{hom})$ in \eqs{eq:lin}.  To our
knowledge, however, there are no studies of this dependency in the
case of monolayers. For bulk colloids, plenty of works have shown that
the specific interparticle forces will affect this dependency, also
whether $\Gamma$ decreases or increases with $\phi$ (see, e.g.,
\rcites{ToOp94,HaIc95,MLP10,Piaz14}). Therefore, it is difficult to
advance a conjecture on the behavior of the function $\Gamma(\phi)$
for a monolayer. Here, we simply assume (without convincing
justification, however) that the variation of $\Gamma$ with $\phi$ is
less than an order of magnitude, as is the case in bulk colloids.
Thus, one can take the value of $\Gamma$ for an isolated particle
(dilute limit) for reference purposes: this can be estimated with
Stoke's drag formula for no--slip boundary conditions as $\Gamma
\approx 1/(6\pi\eta R)$.
Finally, the diffusion coefficient $D_0$ will also depend on $\phi$
according to the interparticle forces; this dependency is studied in
App.~\ref{app:D0} for several realistic models. For reference
purposes, we quote its value for an ideal gas at room temperature,
$D_\mathrm{ideal} 
\approx 0.04\Gamma\times 10^{-19} \textrm{ J}$, see \eq{eq:Dideal}.
However, unlike $\Gamma$, the coefficient $D_0$ can change with $\phi$
by orders of magnitude (see \fig{fig:D0}).

With these choices for the parameter values, the relevant length
scales of the model are estimated as
\begin{subequations}
\begin{equation}
  \label{eq:numLh}
  \frac{L_\mathrm{hydro}}{\mu\mathrm{m}} \approx 
  \frac{2}{3\phi} \frac{R}{\mu\mathrm{m}} ,
\end{equation}
\begin{equation}
  \frac{L_\mathrm{cross}}{\mu\mathrm{m}} \approx \frac{3\times 10^6}{\phi} 
  \frac{D_\mathrm{ideal}}{D_0(\phi)} \left(\frac{R}{\mu\mathrm{m}}\right)^2 ,
\end{equation}
and the relevant time scales are
\begin{equation}
  \frac{\tau_\omega}{\mathrm{s}} \approx \frac{0.4\times 10^{-6}}{\phi^2}
  \left(\frac{R}{\mu\mathrm{m}}\right)^2 ,
\end{equation}
\begin{equation}
  \frac{\tau_2(k=1/L_\mathrm{cross})}{\mathrm{s}} \approx \frac{6\times 10^6}{\phi^2}
  \left(\frac{D_\mathrm{ideal}}{D_0(\phi)}\right)^2 
  \left(\frac{R}{\mu\mathrm{m}}\right)^4 .  
\end{equation}
\end{subequations}
Several observations are in order:
\\
(i) The ratio 
  \begin{equation}
    \frac{L_\mathrm{cross}}{L_\mathrm{hydro}} \approx \frac{5\times 10^6}{\phi} 
    \frac{D_\mathrm{ideal}}{D_0(\phi)} \frac{R}{\mu\mathrm{m}} ,
  \end{equation}
  will be very large in any case, also for the smallest colloidal
  particles, $R\sim 1\;\mathrm{nm}$. Therefore, the separation of
  scales, $L_\mathrm{hydro} \ll L_\mathrm{cross}$, assumed in the
  theoretical analysis holds under realistic conditions.
  \\
  (ii) The length scale $L_\mathrm{hydro}$ is much larger than the
  mean interparticle separation, $\ell \sim R/\sqrt{\phi}$, only for a
  dilute system. A dense system has $L_\mathrm{hydro} \sim \ell$, and
  therefore the regime of normal collective diffusion, $k^{-1}\ll
  L_\mathrm{hydro}$, would be masked in such case presumably by
  single--particle effects.
  \\
  (iii) The time scale for relaxation of the velocity of a single
  particle, on which the overdamped approximation~(\ref{eq:vfield}) is
  based, can be estimated as $\tau_\mathrm{relax} \sim
  \varrho_\mathrm{fluid} R^2/\eta$ (see, e.g., \cite{Dhon96}), so that
  $\tau_\omega/\tau_\mathrm{relax}\sim (L_\mathrm{hydro}/R)^2\sim
  \phi^{-2}$. 
  Therefore, $\tau_\omega > \tau_\mathrm{relax}$ and, in view of
  \eqs{eq:taularge}, the relevant time scales are consistent with the
  approximation of overdamped motion.
  \\
  The anomalous diffusion predicted in the intermediate regime
  $L_\mathrm{hydro} \ll k^{-1} \ll L_\mathrm{cross}$ has been observed
  experimentally: the reinterpretation of old experimental data
  obtained through dynamic light scattering \cite{LRW95} allows one to
  conclude that the coefficient of collective diffusion diverges
  \cite{LCXZ14}. The monolayer of this experiment can be modelled as a
  collection of hard disks of radius $R\approx 120\;\mathrm{nm}$, the
  probed values of the packing fraction being in the range
  $\phi\approx 0.12 - 0.57$.  The above estimates give the range
  $L_\mathrm{hydro} \approx 0.1-0.7\;\mu\mathrm{m}$, $L_\mathrm{cross}
  \approx 6-200\;\mathrm{mm}$, $\tau_2 (k=1/L_\mathrm{cross})\approx
  10 - 10^5\;\mathrm{s}$ for this experimental configuration, so that
  the crossover to the regime of time--dependent hydrodynamic
  interactions is technically unobservable.  However, it is possible
  to reduce the values of $L_\mathrm{cross}$ and $\tau_2$ by
  considering smaller particles or a denser monolayer. For instance, a
  monolayer formed by charged nanoparticles with $R\sim 10
  \;\mathrm{nm}$, at high densities ($\phi\sim 1$) such that one could
  assume $D_0/D_\mathrm{ideal}\sim 10^2$ (see \fig{fig:D0}), is
  predicted to have $L_\mathrm{hydro} \sim 7\; \mathrm{nm}$,
  $L_\mathrm{cross} \sim 3\; \mu\mathrm{m}$ and $\tau_2
  (k=1/L_\mathrm{cross})\sim 6\;\mu\mathrm{s}$. Therefore, the
  observation of the crossover and of the signature by the
  time--dependent hydrodynamic
  interactions 
  on the intermediate scattering function, see \fig{fig:greenktstab},
  should be within reach of up-to-date experimental techniques.
  Actually, given that $L_\mathrm{cross}$ can have values in the
  micrometer range, the effect of ambient vortex relaxation on the
  collective dynamics could be relevant, beyond light scattering
  observations, for the macroscopic rheological properties of the
  monolayer.

\section{Conclusions}
\label{sec:end}

We have generalized the result on anomalous diffusion in colloidal
monolayers derived in \rcite{BDGH14} by including the relaxational
dynamics of the vorticity in the ambient fluid. In this expanded
framework, the result predicted in \rcite{BDGH14} and observed
experimentally in \rcite{LRW95,LCXZ14} must be understood as an
intermediate asymptotic behavior. Two well separated length scales,
$L_\mathrm{hydro}$ and $L_\mathrm{cross}$, have been identified that
separate three dynamical regimes in the evolution of the monolayer: on
spatial scales well below $L_\mathrm{hydro}$, the effect of the
hydrodynamic interactions mediated by the ambient fluid is negligible.
On scales between $L_\mathrm{hydro}$ and $L_\mathrm{cross}$, the
collective dynamics of the monolayer is dominated by the hydrodynamic
interactions as if instantaneous, and the model of \rcite{BDGH14} is
recovered. Finally, on scales well above $L_\mathrm{cross}$ a novel
dynamical regime is predicted where the time dependency of the
hydrodynamic interactions is relevant. We have discussed how each
regime would appear in the experimentally accessible intermediate
scattering function. We have provided a detailed discussion of the
influence of the system parameters on the values of the lengths
$L_\mathrm{hydro}$ and $L_\mathrm{cross}$, and of the possible
observation of the features specific to the different dynamical
regimes.

\appendix

\section{Analytic structure of $\hat{G}(k,s)$}
\label{sec:complexG}

\begin{figure}
  \hfill\epsfig{file=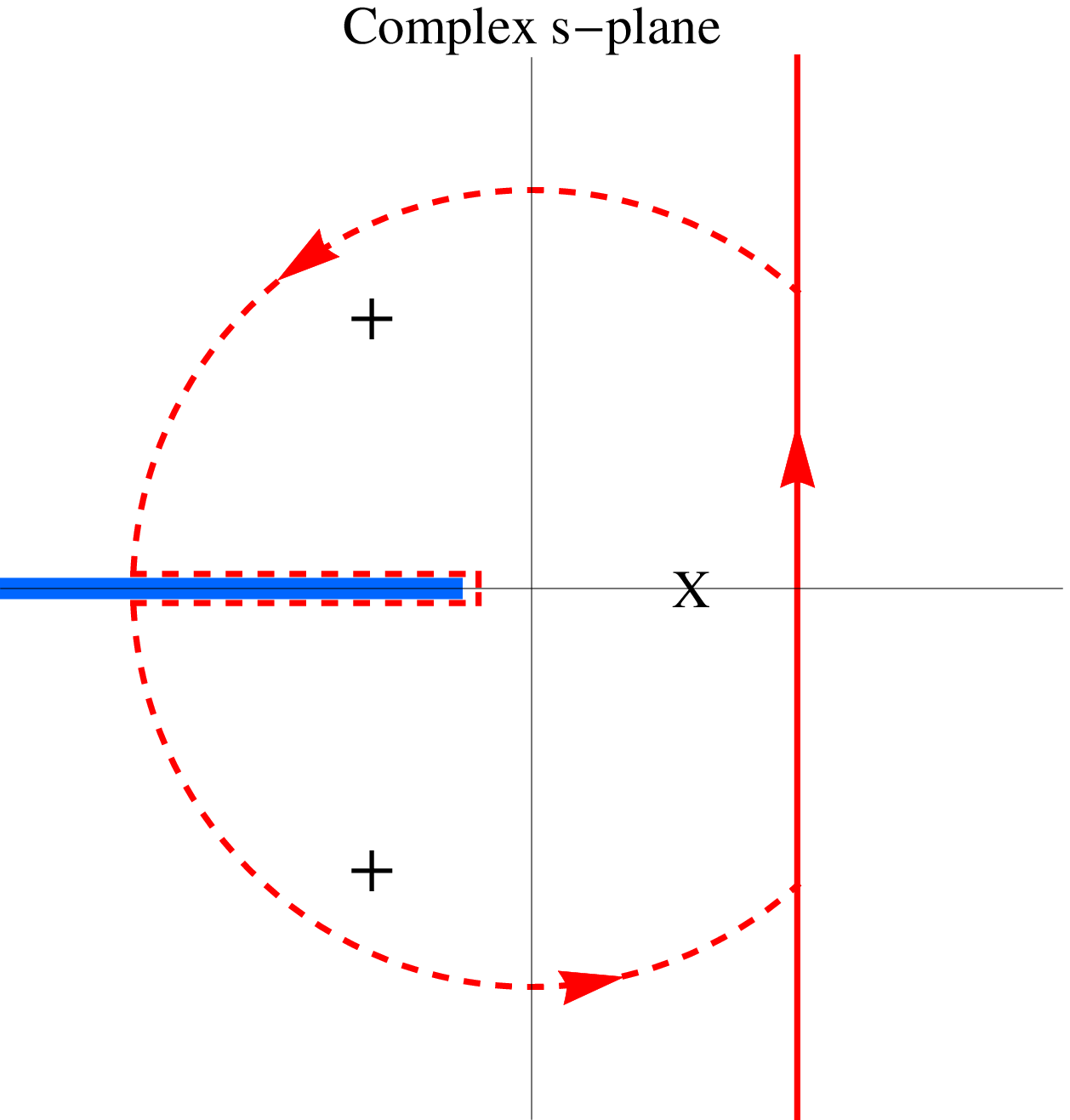,width=.3\textwidth}
  \hfill\epsfig{file=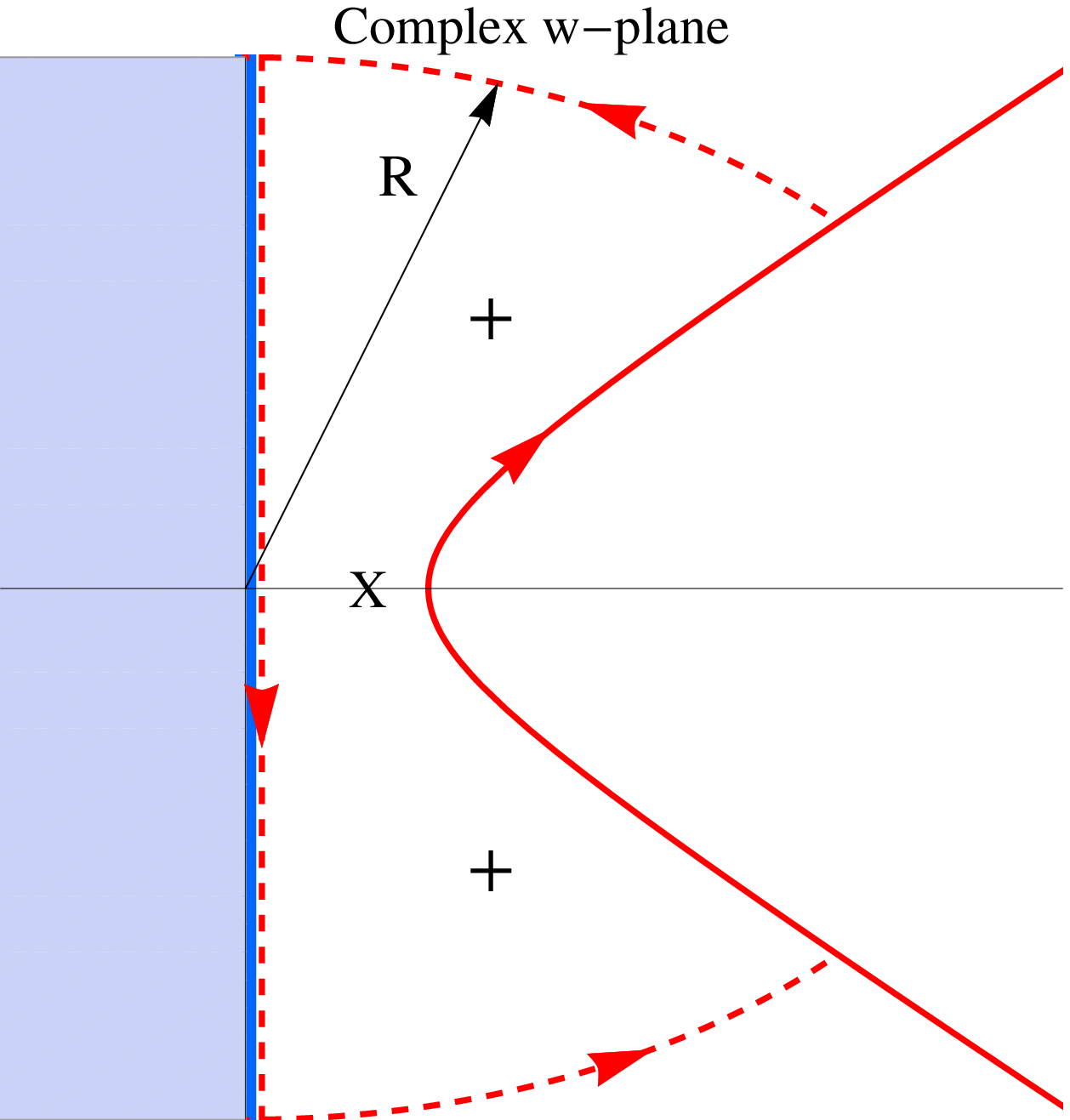,width=.3\textwidth}
  \hspace*{\fill}
  \caption{Analytical structure of $\hat{G}(k,s)$. The complex
    $s$--plane (left) and the complex $w$--plane (right) are related
    by the conformal transformation~(\ref{eq:w}). The thick blue line in
    the $s$--plane is the branch cut, which is transformed into the
    excluded region $\mathrm{Re}\;w<0$ in the $w$--plane. The symbols
    indicate the relative positions of the singularities of
    $\hat{G}(k,s)$: ``+'' when $D_0>0$ and $k<k_c$, ``X'' when $D_0<0$
    or when $D_0>0$ and $k_c<k$. The red line is the integration path
    in the Mellin formula (\eq{eq:mellin} or \eq{eq:mellin2}); the
    dashed red lines correspond to the paths $K_R$ and $L_R$ (see
    \eq{eq:mellin3}) completing the original path into a closed
    contour.
      \label{fig:complexG}}
\end{figure}

\begin{figure}
  \hfill\epsfig{file=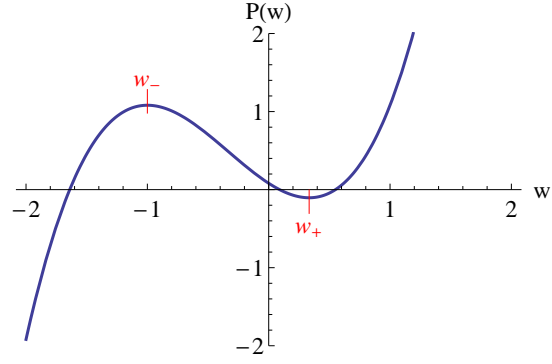,width=.4\textwidth}
  \hspace*{\fill}
  \caption{Graph of the polynomial $P(w)$ defined by \eq{eq:polyP} for generic
    values of the parameters $\lambda>0$ and $k$. Note that changes in
    $k$ amount just to a vertical shift of the graph. The positions of
    the local extrema, $w_\pm$, are given by \eq{eq:wpm}.
    \label{fig:poly}}
\end{figure}

\begin{figure}
  \hfill\epsfig{file=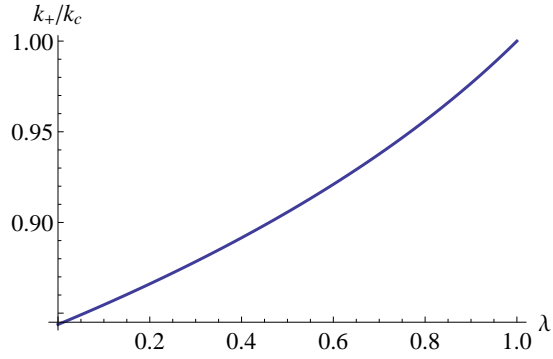,width=.4\textwidth}
  \hspace*{\fill}
  \caption{Plot of $k_+/k_c$ as function of $\lambda>0$, where $k_+$ is
    given by the condition $P(w_+)=0$.
    \label{fig:kplus}}
\end{figure}

The presence of a square root in the definition of the function
$\hat{G}(k,s)$, see Eqs.~(\ref{eq:H}, \ref{eq:green}), implies that it
is defined in the complex $s$--plane with a branching point, see
\eq{eq:sbranch}, and cut along the negative real axis between the
branching point and $-\infty$, see \fig{fig:complexG}. It is useful
to define a conformal transformation to a new complex variable as
\begin{equation}
  \label{eq:w}
  w = \sqrt{1-\frac{s}{s_\mathrm{branch}}} ,
\end{equation}
that maps the cut $s$--plane into the half--plane $\mathrm{Re}\; w >
0$. We introduce the auxiliary quantities
\begin{subequations}
  \label{eq:aux}
\begin{equation}
  \sigma := \mathrm{sign}\; D_0 ,
\end{equation}
\begin{equation}
  \lambda := \sigma \frac{L_\mathrm{hydro}}{L_\mathrm{cross}} 
  = \frac{\tau_\omega D_0}{L_\mathrm{hydro}^2} ,
\end{equation}
\begin{equation}
  k_c := \frac{2\sigma}{L_\mathrm{cross}} (1-\lambda)^{-1} ,
\end{equation}
\end{subequations}
in terms of the length scales defined in Eqs.~(\ref{eq:Lhydro},
\ref{eq:Lcross}), so that the Mellin formula~(\ref{eq:mellin}) becomes
\begin{subequations}
  \label{eq:mellin2}
  \begin{equation}
    G (k, t>0) = \int_{\mellin'}
    dw \; \xi (w,t) ,
  \end{equation}
  with the integrand 
  \begin{equation}
    \xi(w,t) := \frac{w (1+w) \mathrm{e}^{(1-w^2) t s_\mathrm{branch}}}{i \pi P(w)} ,
  \end{equation}
  \begin{equation}
    \label{eq:polyP}
    P(w) := w^3 + w^2 - (1-\lambda) w + (1-\lambda) \left(\frac{k_c}{k} - 1 \right) ,
  \end{equation}
  and the transformed integration path
  \begin{equation}
    \mellin' := \left\{ \mathrm{Re} \; (w^2-1) = p' > 0 \right\} ,
  \end{equation}
\end{subequations}
i.e., the branch of the hyperbole $(\mathrm{Re}\; w)^2 -
(\mathrm{Im}\; w)^2 = 1 + p'$ in the half--plane $\mathrm{Re}\; w > 0$
and to the right of any singularity of the integrand, see
\fig{fig:complexG}. These singularities are given by the roots of the
polynomial $P(w)$ with non-negative real part (possibly barring the
non-generic cases that $w=0$ or $w=-1$ are roots). In order to analyze
the character of the roots, we assume
$L_\mathrm{hydro}<L_\mathrm{cross}$, so that $-1<\lambda< +1$ and
$1-\lambda > 0$ (actually, as argued in Sec.~\ref{sec:exp}, the
physically relevant situations correspond to $|\lambda|\ll 1$).
Several cases can be distinguished:
\\
\textbf{(1)} Either $\sigma=-1$, or $\sigma=+1$ and $k>k_c$: although
$k_c$ can have any sign, it is $(1-\lambda)(k_c/k-1) < 0$.  Therefore,
application of Descartes' rule of signs 
to the polynomial $P(w)$ allows one to conclude that there is always a
positive real root. The other two roots are either negative or complex
conjugate of each other. In any case, the sum of the three roots must
equal $-1$ (minus the coefficient of $w^2$ in $P(w)$), so that their
real parts must be negative and they are thus irrelevant. In
conclusion, the only relevant singularity in the $w$--plane is a
simple pole on the positive real axis.
\\
\textbf{(2)} $\sigma=+1$ and $k<k_c$: now $k_c>0$ and
$(1-\lambda)(k_c/k-1) > 0$, and Descartes' rule of signs leads to the
conclusion that there is one negative, and thus irrelevant root. The
other two roots are either positive or complex conjugate of each
other. To clarify this issue, consider the positions of the local
extrema of $P(w)$, given as
\begin{equation}
  \label{eq:wpm}
  P'(w)=0 \Rightarrow w_\pm = \frac{1}{3} \left[ - 1 \pm \sqrt{1 + 3 (1-\lambda)} \right] ,
\end{equation}
with $w_-<0$ and $w_+>0$. Qualitatively, $P(w)$ has the graph shown in
\fig{fig:poly}, so that two positive roots can only exist provided
$P(w_+)<0$. One defines $k_+$ by the condition $P(w_+)=0$; although an
explicit expression for $k_+$ as a function of $\lambda$ can be
written down, it suffices with the plot in \fig{fig:kplus}, showing that,
for all practical purposes, one can assume $k_+ \approx k_c$.
Therefore, ignoring the narrow range $k_+ < k < k_c$ where the
polynomial $P(w)$ has two positive roots,
the roots will be complex conjugate of each other. One can further
show that, in such case, their real part is positive: denoting the
roots as $w_1$, $w_2$, $w_3$, with $w_1<0$, $w_3^* = w_2$, one can
inspect the coefficient of $w$ in $P(w)$:
\begin{equation}
  w_1 w_2 + w_1 w_3 + w_2 w_3 = 2 w_1 \mathrm{Re}\; w_2 + |w_2|^2 
  = - (1-\lambda) < 0 ,
\end{equation}
so that $\mathrm{Re}\; w_2 > 0$ necessarily. In conclusion, the
relevant singularities are two complex conjugate simple poles in the
half--plane $\mathrm{Re}\;w > 0$.

The pole structure of $\hat{G}(k,s)$ is summarized \fig{fig:complexG}.
The Mellin formula in \eq{eq:mellin2} can be written in terms of
explicit real Riemann integrals by application of the theorem of the
residues: one builds a closed contour as indicated in \fig{fig:complexG}
in the limit $R\to \infty$, consisting of a straight line along the
(transformed) branch cut ($L_R$ at $\mathrm{Re}\; w = 0^+$) and two arcs
of circle (of radius $R$ and denoted $K_R$) joining the straight line
with the hyperbole $\mellin'$. Then, \eq{eq:mellin2} can be written as
\begin{equation}
  \label{eq:mellin3}
  G(k,t) = 
  2 \pi i \sum_n \mathrm{Res}_{w_n} \xi (w,t) 
  - \lim_{R\to\infty} \int_{K_R \cup L_R} dw\; \xi(w,t) .
\end{equation}
The residues contribute either a real exponential when $D_0<0$ or when
$D_0>0$ and $k_c<k$ (case 1 above), or exponentially damped
oscillations when $D_0>0$ and $k_c<k$ (case 2).
The integral along the arcs $K_R$ vanishes in the limit $R\to\infty$,
as follows from a standard application of Jordan's lemma (which is
more easily done in terms of the original variable $s$ in the Mellin
formula in \eq{eq:mellin}). 
The integral along the straight line, $L_R := \left\{ w = i y \; | \;
  |y| < R \right\}$ in the sense of decreasing $y$, can be simplified
after some algebraic manipulations and a change to the new integration
variable $x=y^2$:
\begin{subequations}
  \label{eq:integralbranch}
  \begin{equation}
    \lim_{R\to\infty} \int_{L_R} dw \; \xi(w,t)
    = \frac{1-\lambda}{\pi} \frac{k_c}{k} \int_0^{\infty} dx \; 
    \frac{\sqrt{x} \; \mathrm{e}^{(1+x) t s_\mathrm{branch}}}{|P(i \sqrt{x})|^2} ,
  \end{equation}
  where
  \begin{eqnarray}
    |P(i \sqrt{x})|^2 & = & x^3 + (3-2\lambda) x^2 
    + (1-\lambda) \left( 3-\lambda-\frac{2 k_c}{k} \right) x 
    \nonumber \\
    & & \mbox{} + (1-\lambda)^2 \left(\frac{k_c}{k} - 1 \right)^2 .
  \end{eqnarray}
\end{subequations}

\section{Green function in the limit $k\to 0$}
\label{sec:smallk}

Here \eqs{eq:greensmallk} are derived. In the limit $k\to
0$, the roots of $P(w)$ are given by the approximate equation (see
\eq{eq:polyP})
\begin{equation}
  P(w) \approx w^3 + (1-\lambda)\frac{k_c}{k} = 0 .
\end{equation}
Since $1-\lambda > 0$, the solutions of this equation in the
half--plane $\mathrm{Re}\; w>0$ depend on the sign of $D_0$ (through
the sign of $k_c$, see \eqs{eq:aux}):
\begin{equation}
  w = \left[ (1-\lambda) \frac{|k_c|}{k} \right]^{1/3}
  \times \left\{
    \begin{array}[c]{cl}
      1 & \mathrm{if}\; D_0 < 0 , \\
      & \\
      \mathrm{e}^{\pm i \pi/3} & 
      \mathrm{if}\; D_0 > 0 .      
    \end{array}
    \right.
\end{equation}
Therefore, from Eqs.~(\ref{eq:sbranch}, \ref{eq:w}) one deduces the
position of the singularities of $\hat{G}(k,s)$ in the complex
$s$--plane as $k\to 0$:
\begin{equation}
  \label{eq:polesk0}
  s = \frac{1}{\tau_2(k)} \times \left\{
    \begin{array}[c]{cl}
      1 & \mathrm{if}\; D_0 < 0 , \\
      & \\
      \mathrm{e}^{\pm i 2 \pi/3} & 
      \mathrm{if}\; D_0 > 0 ,      
    \end{array}
    \right.
\end{equation}
in terms of the time scale $\tau_2(k)$ defined by \eq{eq:tau2}.

The inversion of the approximated Green
function~(\ref{eq:greentimeHI}) with the Mellin
formula~(\ref{eq:mellin}) is facilitated by performing the change of
variable $z = \sqrt{s \tau_2(k)}$. This is nothing else but the
conformal transformation~(\ref{eq:w}) in the limit $k\to 0$ with an
appropriate rescaling so that the singularities have a finite position
in this limit, see \eq{eq:polesk0}. With this new variable,
Eqs.~(\ref{eq:mellin}, \ref{eq:greentimeHI}) lead to
\eq{eq:greensmallkscaling} where
\begin{subequations}
  \label{eq:gammadef}
  \begin{equation}
    \gamma(u) := \frac{1}{i \pi} \int_{\hat\mellin} dz\; 
    \frac{z^2}{z^3 + \sigma} \mathrm{e}^{u z^2} ,
  \end{equation}
  with the transformed integration path
  \begin{equation}
    \label{eq:hatpath}
    \hat{\mellin} := \{ \mathrm{Re}\; z^2 = \hat{p} > 0 \},
  \end{equation}
\end{subequations}
i.e., the branch of the hyperbole $(\mathrm{Re}\; z)^2 -
(\mathrm{Im}\; z)^2 = \hat{p}$ in the half--plane $\mathrm{Re}\; z >
0$ and to the right of any singularity.

The expresion of $\gamma(u)$ in \eq{eq:gamma} follows from the
evaluation of the integral~(\ref{eq:gammadef}) by closing the contour
$\hat{\mellin}$ as in \eq{eq:mellin3} (see \fig{fig:complexG}). This
gives a contribution of the poles and a contribution $I(u)$, see
\eq{eq:I}, of the integral along the (transformed) branch cut
$\mathrm{Re}\; z = 0^+$. Of course, these expressions could have been
obtained alternatively by taking the limit $k\to 0$ in the exact
expressions~(\ref{eq:mellin3}, \ref{eq:integralbranch}).

\section{Properties of the function $\gamma(u)$}
\label{sec:greenkt}

The function $\gamma(u)$ defined by \eq{eq:gammadef} can be expressed
as in \eq{eq:gamma}. Since the integrals
\begin{equation}
  \frac{d^n I}{d u^n} = \frac{1}{\pi} \int_0^\infty dx\;
  \frac{(-x)^n \sqrt{x}}{x^3+1} \mathrm{e}^{-u x}
\end{equation}
converge for any $n$ if $u>0$, the function $\gamma(u)$ and its
derivatives exist for any $u>0$. To obtain the behavior as $u\to 0$,
one writes the function $I(u)$ as follows:
\begin{eqnarray}
  I(u) & = & \frac{1}{\pi} \int_0^\infty dx\; \frac{\sqrt{x}}{1+x^3} ( 1 - u x )
  + \frac{1}{\pi} \int_0^\infty dx\; \frac{\sqrt{x}}{1+x^3} 
  \left( \mathrm{e}^{-ux} - 1 + u x \right)
  \nonumber \\
  & = & \frac{1-2 u}{3} + \frac{u^{3/2}}{\pi} \int_0^\infty d\xi\; \frac{\sqrt{\xi}}{u^3+\xi^3} 
  \left( \mathrm{e}^{-\xi} - 1 + \xi\right) 
  \nonumber \\
  & \stackrel{\tiny u\to 0}{\sim} & \frac{1-2 u}{3} + \frac{u^{3/2}}{\pi} \int_0^\infty d\xi\; \frac{\sqrt{\xi}}{\xi^3} 
  \left( \mathrm{e}^{-\xi} - 1 + \xi \right) 
  \nonumber \\
  & \sim & \frac{1}{3} \left( 1 - 2 u + \frac{4}{\sqrt{\pi}} u^{3/2} \right) .
\end{eqnarray}
Inserting this result in \eq{eq:gamma} and expanding further in $u\to
0$, one arrives at \eq{eq:smallu}. 
Actually, one can show that $\gamma(u)$ has an expansion in powers of
$u^{3/2}$ about $u=0$: changing the integration variable to $\zeta = z
\sqrt{u}$ in the definition~(\ref{eq:gammadef}), one has
\begin{equation}
  \gamma(u) = \frac{1}{i\pi} \int_{\mathfrak{C}} d\zeta\; 
  \frac{\zeta^2 \mathrm{e}^{-\zeta^2}}{\zeta^3+\sigma u^{3/2}} ,
  \quad
  \mathfrak{C} = \{ \mathrm{Re}\; \zeta^2 = 1 
  \} ,
\end{equation}
where the integration path $\mathfrak{C}$ can be taken to be
$u$--independent because the singularities of the new integrand
approach zero as $u\to 0$. Since $|\zeta|\neq 0$ along the
integration path, one can Taylor--expand the integrand uniformly as
$u\to 0$ and obtain
\begin{subequations}
\label{eq:expandu}
\begin{displaymath}
  \gamma(u)
  = \frac{1}{i\pi} \int_{\mathfrak{C}} d\zeta\; 
  \frac{\mathrm{e}^{-\zeta^2}}{\zeta}
  \sum_{n=0}^{\infty} \left(-\sigma \frac{u^{3/2}}{\zeta^3}\right)^n
  =  \sum_{n=0}^{\infty} a_n u^{3 n/2} ,
\end{displaymath}
\begin{equation}
  a_n := \frac{(-\sigma)^n}{i\pi} \int_{\mathfrak{C}} d\zeta\; 
  \frac{\mathrm{e}^{-\zeta^2}}{\zeta^{1+3 n}} .
\end{equation}
\end{subequations}

In the opposite limit, $u\to +\infty$, one notices that the function
$I(u)$ is formally analogous to a Laplace transform, see \eq{eq:I},
whose asymptotic behavior can be evaluated with Laplace's method
\cite{BeOr78}:
\begin{equation}
  \label{eq:Ilargeu}
  I(u) \sim \frac{1}{\pi} \int_0^{\infty} dx\; \sqrt{x}\,
  \mathrm{e}^{- u x}
  = \frac{u^{-3/2}}{2\sqrt{\pi}} .
\end{equation}
Therefore, this algebraic decay dominates over the exponential decay
when $\mathrm{sign}\;D_0=+1$, but is subdominant compared to the
exponential growth when $\mathrm{sign}\;D_0=-1$. In this manner,
\eq{eq:largeu} is obtained.

\section{The coefficient $D_0$ in different fluid
  models}
\label{app:D0}

By using thermodynamic identities, \eq{eq:D0} can be written as
\begin{equation}
  \label{eq:D0comp}
  D_0 = \Gamma 
  \left.\frac{\partial p}{\partial \varrho}\right)_T (\varrho=\varrho_\mathrm{hom}) ,
\end{equation}
in terms of the equation of state $p(\varrho, T)$ for the 2D pressure
of the colloidal monolayer. In the dilute limit, the ideal gas
approximation provides 
\begin{equation}
  \label{eq:Dideal}
  D_0 = D_\mathrm{ideal} = \Gamma T 
\end{equation}
(with the temperature $T$ given in units of energy).

\begin{figure}
  \hfill\epsfig{file=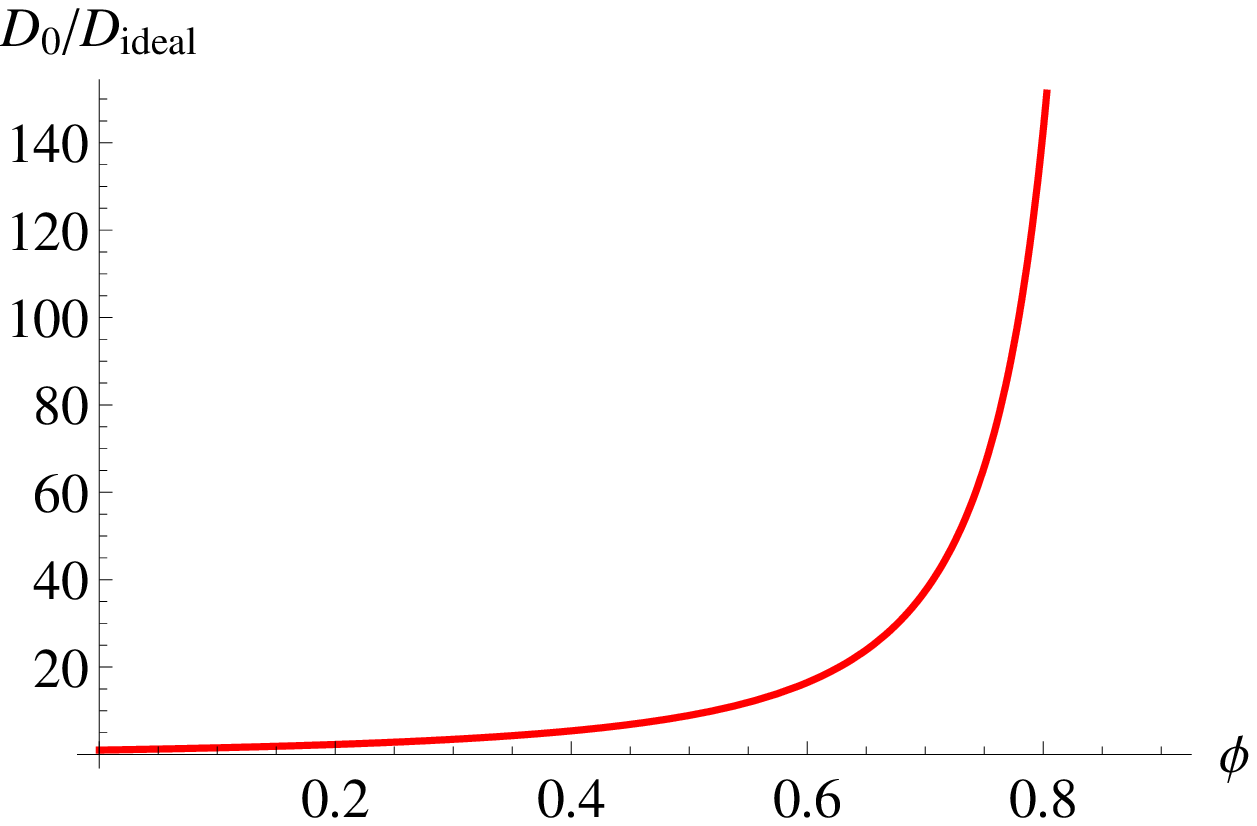,width=.4\textwidth}
  \hfill\epsfig{file=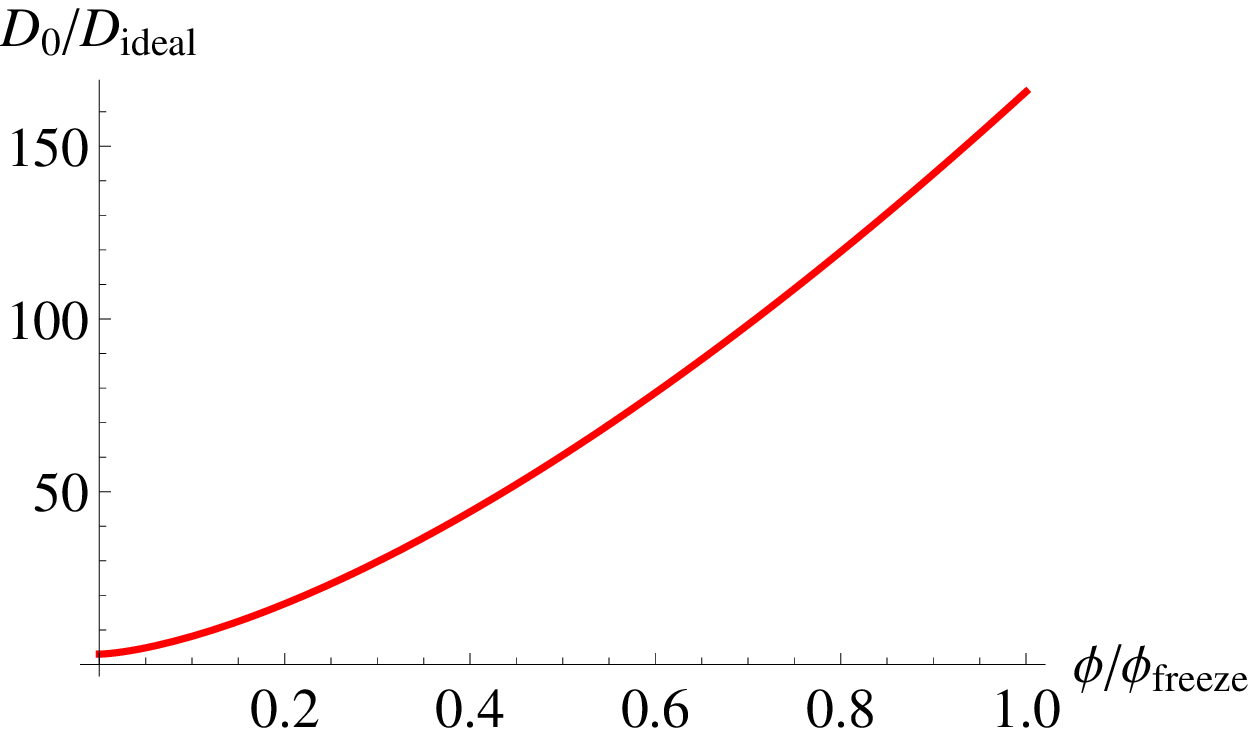,width=.4\textwidth}
  \hspace*{\fill}
  \caption{(\textit{Left}) The ratio $D_0/D_\mathrm{ideal}$ for a 2D
    fluid of hard disks, see \eq{eq:Dhard}, as a function of the packing fraction $\phi$
    (which is $\approx 0.91$ at close packing). (\textit{Right}) The
    ratio $D_0/D_\mathrm{ideal}$ for a 2D fluid with \eq{eq:vsoft} as
    interparticle potential, as a function of
    $\phi/\phi_\mathrm{freeze}$, where $\phi_\mathrm{freeze} \approx
    14.45 (R/\zeta)^2$ is the packing fraction at the freezing
    transition.
    \label{fig:D0}}
\end{figure}

If the monolayer can be modelled as a collection of hard disks of radius
$R$, its equation of state can be approximated by the expression
\cite{GZB97}
\begin{equation}
  p = \varrho T \frac{\varrho_c+\varrho}{\varrho_c-\varrho} ,
\end{equation}
with $\varrho_c = (2\sqrt{3} R^2)^{-1}$ the number density for close
packing of disks. From here one gets the ratio
\begin{equation}
  \label{eq:Dhard}
  \frac{D_0}{D_\mathrm{ideal}} = 
  \frac{2}{(1-\phi/\phi_\mathrm{max})^2} - 1 .
\end{equation}
in terms of the maximum packing fraction, $\phi_\mathrm{max}\approx
0.91$, assuming a $\phi$--independent value of the mobility $\Gamma$
(see the discussion in Sec.~\ref{sec:exp}). This function is plotted
in \fig{fig:D0}.

Another realistic model of the monolayer is as a collection of
particles with a soft repulsion, described by the potential
\begin{equation}
  \label{eq:vsoft}
  V(r) = T \left(\frac{\zeta}{r}\right)^3 .
\end{equation}
Here, $\zeta$ is the associated Bjerrum length, which must be
substantially larger than $R$ so that the interparticle repulsion is
indeed dominated by \eq{eq:vsoft} rather than by hard--core effects.
This potential describes the large--separation dominant part of the
electrostatic repulsion between charged particles located at the
interface between a dielectric fluid and an electrolytic solution
\cite{Hurd85,ACNP00,DFO08,DaKr13b} and between polarizable particles in an
external electric field \cite{ASJN08,DaKr13a}, and also the repulsion between
superparamagnetic particles in an external magnetic field
\cite{ZMM97}. The equation of state associated to this potential can
be computed using Montecarlo simulations \cite{DOD10} and the results
for the ratio $D_0/D_\mathrm{ideal}$ are plotted in \fig{fig:D0}
(assuming again a $\phi$--independent mobility).

%

\end{document}